\documentclass[%
mathleft,%
]{an}
\usepackage{natbib}
\usepackage{graphicx}
\usepackage{times}
\usepackage{hyperref}
\usepackage{amssymb}
\begin{document}

% The following seven commands are intended for editorial usage and
% should be ignored by the author(s).
\Pagespan{1}{}% Document's page range. 
% If second parameter is left empty, the last page is computed
% automatically.
\Yearpublication{2015}%
\Yearsubmission{2015}%
\Month{10}%   
\Volume{999}%  
\Issue{92}% 
% \DOI{This.is/not.aDOI}% 

\title{An Evolutionary Sequence of Young Radio Galaxies}

\author{J.~D.~Collier\inst{1,2}\thanks{E-mail: j.collier@uws.edu.au}, %can use \email{j.collier@uws.edu.au}
R.~P.~Norris\inst{1,2},
M.~D.~Filipovi\'c\inst{1},
N.~F.~H.~Tothill\inst{1}}

\titlerunning{An Evolutionary Sequence of Young RGs}
\authorrunning{J.~D.~Collier et al. 2015}
\institute{University of Western Sydney, Locked Bag 1797, Penrith, NSW, 2751, Australia
\and
CSIRO Astronomy and Space Science, Marsfield, NSW, 2122, Australia}

\received{XXXX}
\accepted{XXXX}
\publonline{XXXX}

 %galaxies: formation, 

\abstract{
We have observed the faintest sample of Gigahertz Peaked Spectrum (GPS) and Compact Steep Spectrum (CSS) sources to date, using the Australia Telescope Compact Array. We test the hypothesis that GPS and CSS sources are the youngest radio galaxies, place them into an evolutionary sequence along with a number of other young Active Galactic Nuclei (AGN) candidates, and search for evidence of the evolving accretion mode and its relationship to star formation. GPS/CSS sources have very small radio jets that have been recently launched from the central Supermassive Black Hole and grow in linear size as they evolve, which means that the linear size of the jets is an excellent indicator of the evolutionary stage of the AGN. We use high-resolution radio observations to determine the linear size of GPS/CSS sources, resolve their jets and observe their small-scale morphologies. We combine this with other multi-wavelength age indicators, including the spectral age, colours, optical spectra and Spectral Energy Distribution of the host galaxy, in an attempt to assemble all age indicators into a self-consistent model. We observe the most compact sources with Very Large Baseline Interferometry, which reveals their parsec-scale structures, giving us a range of source sizes and allowing us to test what fraction of GPS/CSS sources are young and evolving.}

\maketitle

\section{Introduction}

\subsection{AGN evolution}

\label{intro}

It has long been suggested that the Active Galactic Nuclei (AGN) and star formation (SF) within a galaxy are closely related and play a significant role in the evolution of galaxies. A typical scenario \citep[e.g.][]{Best2006, Croton2006, Hardcastle2007} involves the merger of a gas-rich spiral with another spiral or elliptical, causing an intense burst of SF together with ``quasar-mode accretion'' onto a central supermassive black hole (SMBH). The SMBH grows both through coalescence and through the accretion of cold disc gas from the host galaxies, producing a rapid growth of the SMBH mass. The outflows driven by the resulting powerful quasar winds quench SF activity \citep{2007MNRAS.382.1394M,2008MNRAS.389.1750A,2008ApJS..175..356H}, and clear the central regions of fuel, starving the AGN, which begins accreting hot gas inefficiently (``radio mode accretion'').

\cite{2015ApJ...806..147C} find evidence that mergers are involved in triggering radio-loud AGN at all redshifts. \citet{2012MNRAS.423...59S} present the timescales over which the starburst and AGN are triggered for a sample of dust lane early-type galaxies, including the delay of $100-150$ Myr between these two phases. However, these timescales are not known in general, nor whether the starburst always precedes the AGN. 

A dramatic example of this process can be found in the Ultra-Luminous IR Galaxy (ULIRG) F00183-7112 \citep{Norris12}, in which a starbursting merger hosts a hidden powerful radio-loud AGN, whose jets have not yet broken through the shroud of dust and gas, and so is invisible at optical/NIR wavelengths. F00183 also appears to be in a transitionary stage between Gigahertz Peaked Spectrum (GPS) and Compact Steep Spectrum (CSS), with evidence of a turnover at the lowest frequencies. 

Therefore, in order to gain a complete understanding of AGN evolution, the SF and AGN must be considered simultaneously. Here, we test this general AGN model by studying GPS and CSS sources, as well as Infrared-Faint Radio Sources (IFRSs).

\subsection{GPS and CSS sources}

\begin{figure}
\begin{center}
\includegraphics[width=0.33\textwidth,angle=-90]{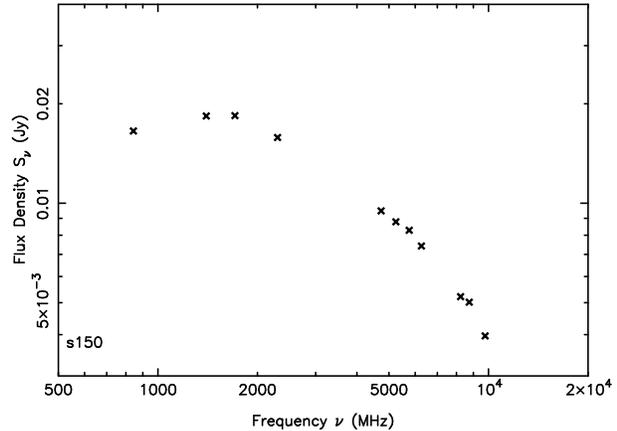}
\caption{\small Radio spectrum of faint GPS source from the ATLAS ELAIS-S1 with component ID S150 (see Section~\ref{ATLAS}), including the SUMSS 843 MHz observations, ATLAS 1.4, 1.7 and 2.3 GHz observations, and the 4.5-10 GHz observations from this project.} 
\label{s150}
\end{center}
\end{figure}

GPS and CSS sources are widely believed to represent the earliest stages of radio-loud AGN evolution \citep[e.g.][]{ODea, 2003PASA...20...69P, 2006A&A...445..889T, 2009AN....330..120F,2011MNRAS.416.1135R,1995A&A...302..317F}. GPS sources turn over at a few GHz (see Fig.~\ref{s150}) and are typically $<$~1~kpc in size. CSS sources turn over between $50-100$ MHz, have steep ($\alpha < -0.8$\footnote{$S\propto\,\nu^{\alpha}$}) spectral indices across the GHz range, and are typically $1-10$ kpc in size. It is generally considered that most GPS sources evolve into CSS sources, which gradually evolve into Fanaroff-Riley Type I (FR~I) and II (FR~II) galaxies \citep{1974MNRAS.167P..31F}. If this is true, then GPS/CSS sources are ideal objects for investigating the birth and early lives of AGN, including any associated SF and its affect on the AGN.

The turnover in the spectrum is generally thought to be caused by synchrotron self absorption (SSA), but could also be due to free-free absorption (FFA) through an inhomogeneous screen \citep[e.g.][]{Tingay}. \citet{2003PASA...20...19M} derives the spectral age ($t$) based on the electron lifetime, which is given by

\begin{equation}
t = 5.03 \times 10^4 \cdot B^{-1.5}[(1+z)\nu_{\rm br}]^{-0.5}~\mbox{years}
\end{equation}

where $B$ is the strength of the magnetic field in milli Gauss (mG) and $\nu_{\rm br}$ is the break frequency in GHz. The break frequency is often difficult to observe, so detailed observations and modelling of the radio spectra are necessary to provide a sufficient estimate of the spectral age. 

Radio imaging not only reveals the spectra of GPS and CSS sources, but also their compact jets, cores, and hot spots. The high-resolution imaging required to reveal such compact details is usually limited to Very Large Baseline Interferometry (VLBI), but can also be achieved with deep high-frequency radio observations. The radio morphologies of GPS and CSS sources divide into symmetric, core-jet and complex, all of which show evidence of compact jets.

At the very earliest stages, the radio power of the jets is thought to grow linearly with size \citep{2000MNRAS.319....8A}, until they reach a size of a few hundred pc, when the jets continue to grow in size but with a roughly constant power \citep{Stas_2008, 2011MNRAS.413.2815S}. \cite{Ross_2015} and \cite{Stas_2008} present dynamical models for the timescales and luminosities of the jet growth beyond 1 kpc, during which the jet travels at supersonic, transonic and subsonic speeds. Therefore, the linear size of the jets is an excellent indicator for the age of the AGN.

\citet{1997AJ....113..148O} present a relation between the projected linear size ($l$) of the radio source and the turnover frequency ($\nu_m$), defined as

\begin{equation}
\label{linear_size_turnover}
\log \nu_m = -0.21(\pm0.05) - 0.65(\pm0.05) \log l~.
\end{equation}

\noindent The minimal scatter around this linear fit shows that the there is a continuous rather than bimodal distribution, which implies that GPS sources are simply scaled-down CSS sources. This is consistent with the great deal of overlap that exists between GPS and CSS sources, which are chosen arbitrarily from their turnover frequencies.

Until recently, our understanding of GPS and CSS sources was limited to very bright (Jy-level) samples. Even still, their properties at faint levels are generally unknown. In particular, no faint CSS sources have been observed, although this may be a selection effect. \citet{2012MNRAS.421.1644R} and \citet{2010MNRAS.408.1187H} have compiled catalogs of mJy-level GPS and CSS sources, although until now it has not been plausible to study their properties at even fainter ($\mu$Jy) levels across a large sample. 

It is generally accepted that GPS and CSS sources are young and evolving radio sources that are developing into large-scale radio sources. Evidence for the young RG hypothesis includes their appearance as scaled-down versions of FR I/II galaxies, proper motion measurements of their hotspot expansion speeds and models of their radio spectra and spectral ages \citep{2009AN....330..120F,2009AN....330..303F,ODea}. However, alternative interpretations for GPS and CSS sources include that they are: (1) frustrated by interactions with dense gas and dust in their environment, which halts the expulsion of the jets and is responsible for the compactness of the sources \citep{1984AJ.....89....5V,1990A&A...232...19B}; (2) prematurely dying radio sources \citep{2009AN....330..303F,2010MNRAS.402.1892O}; (3) recurrent radio galaxies \citep{1990A&A...232...19B,2012A&A...545A..91S}. We aim to directly test these hypotheses over a large, faint sample. The study of GPS and CSS sources at faint levels is likely to give great insight into the genesis and evolution of powerful radio galaxies.

\subsection{IFRSs}

\citet{Collier_2014} and \citet{Andreas_IFRS} show that IFRSs are a new class of high-redshift radio galaxy (HzRG), $\sim$20\% of which are GPS / CSS sources at redshifts between $2 < z < 3$. The first generation of IFRSs found by \cite{ATLAS_CDFS} and \cite{ATLAS_ELAIS}, which were not detected in the IR, are thought to represent HzRGs at $z > 3$, and observations suggest that $>50$\% of these consist of GPS/CSS sources, based on curvature in the radio spectra and compact morphologies. \cite{Norris2011} estimate a sky density of the first generation IFRSs of 7 per deg$^2$, whereas only $\sim$200 HzRGs are currently known at $z > 2$ with $L_{1.4~ \rm GHz} > 10^{26}$ W/Hz. 
Therefore, IFRSs are dominated by GPS/CSS sources at high redshift and there may be as many as $\sim$100 000 of them across the $\mu$Jy sky.

\section{An Evolutionary Sequence of Young Radio Galaxies}

Here we present a progress report on our project to map out an evolutionary sequence of young radio galaxies (RGs), including the general strategy and current status. The results from this project will be presented in a study of our mid-strength sample (Collier et al., 2015, in prep.), faint sample (Collier et al., 2016, in prep.) and VLBI sample (Collier et al., in prep.).

\subsection{Aim of the Project}

We aim to construct an evolutionary sequence for the early stages of AGN development, with a particular focus on faint GPS and CSS sources. The place of GPS/CSS sources within an evolutionary sequence is not well understood. We test the hypothesis that GPS and CSS sources are the youngest RGs, place them into an evolutionary sequence along with a number of other young AGN candidates (including IFRSs), and search for evidence of the evolving accretion mode and its relationship to SF, in order to gain a more complete understanding of AGN evolution (see section~\ref{intro}). We study faint GPS/CSS sources from radio through to X-rays over a large sample size in order to detect meaningful indicators of their age. 

\subsection{Age Indicators}

The indicators we use to derive independent estimates of the AGN age include the:

\begin{itemize}
\item jet sizes
\item spectral age
\item separation of nuclei (where available)
\item colours, Spectral Energy Distribution (SED) and optical spectra of the host galaxy
\end{itemize}

The main age indicator we use is the jet size, as measured from our high resolution radio observations. We use the jet sizes as input for dynamical models such as \cite{Ross_2015} and \cite{Stas_2008}, which give an estimate of the age of the jets. Another age indicator we use is the break frequency, or where this is unavailable, the turnover frequency, obtained from the radio spectrum. We fit power law, SSA and FFA models with and without a break frequency to all available radio flux density measurements, and derive the turnover and break frequencies from the best fit model. Where break frequencies are derived, the spectral age is estimated. The host colours, SEDs and spectra are obtained using all available IR and optical data.

Combining these age indicators enables us to identify if there is any evolutionary correlation between these properties, if they can be assembled into a self-consistent model, and if this is consistent with the currently accepted model.

\subsection{Data}

\subsubsection{Faint Sample: ATLAS}

\label{ATLAS}

We made use of 1.4, 1.7 and 2.3~GHz observations from the Australia Telescope Large Area Survey \citep[ATLAS; Franzen et al., 2015, submitted;][]{ATLAS_CDFS, ATLAS_ELAIS}, which is the widest deep radio survey, covering $\sim$7 square degrees in the {\it Chandra} Deep Field South (CDFS) and the European Large Area ISO Survey South 1 (ELAIS-S1) fields down to an r.m.s. of $\sim$15~$\mu$Jy beam$^{-1}$ at 1.4~GHz. ATLAS overlaps with deep observations in the X-ray, optical and IR, including those from {\it Chandra}, {\it Spitzer}, {\it Herschel} and {\it VIDEO}. The third data release (DR3; Franzen et al. 2015, submitted) contains more than 5 000 galaxies, about half of which are AGN, and most of which have spectroscopic redshifts. Importantly, the ATLAS fields have been observed across many low radio frequencies where CSS sources turn over, including 843 MHz observations from the Sydney University Molonglo Sky Survey \citep[SUMSS;][]{2003MNRAS.342.1117M}, Giant Metrewave Radio Telescope (GMRT) observations at 150, 325 \citep{2009MNRAS.395..269S} and 610 MHz (Intema et al., in prep; Ivison et al., in prep), and upcoming Murchison Widefield Array (MWA) and Australian Square Kilometre Array Pathfinder (ASKAP) observations. The Extended CDFS (ECDFS), which covers 0.25 deg$^2$ of the CDFS, has also been observed at 5.5~GHz \citep{2012MNRAS.426.2342H}.

The multi-wavelength data will give valuable information about the environment, revealing how AGN properties such as size, luminosity and duty cycle depend on environment. For some sources, we will have independent measurements of the star-formation rate (e.g. Herschel and H$\alpha$), which will allow us to study the mechanisms of the co-triggering of the AGN and the SF (see section~\ref{intro}) and the effect the expanding jet has on these, none of which has been done at $z > 0$. ATLAS enables this research to take place on such a faint sample of such size for the first time, which will pave the way for the science that will come out of ASKAP's Evolutionary Map of the Universe \citep[EMU;][]{EMU}.

\subsubsection{Mid-strength Sample: SMC}

The close proximity of the Small Magellanic Cloud (SMC) has resulted in extensive studies of its foreground emission during the past few decades, including deep observations from {\it Spitzer}, {\it XMM-Newton} and {\it Herschel}, and redshift measurements for many background quasars \citep{Kozlowski_2011,Kozlowski_2013}. 

Since radio observations are able to penetrate through the foreground dust and gas, this has resulted in making the SMC a large region of sky ($> 20$ square degrees) that contains thousands of background radio sources at mJy levels, most of which are AGN.

We made use of the \citet{2002MNRAS.335.1085F}, \citet{2011SerAJ.183...95C} and \citet{2011SerAJ.182...43W} mosaics that were compiled by merging many SMC observations together at 0.84, 1.35, 2.37, 4.80 and 8.64~GHz, which reach r.m.s. values of $0.4-0.8$ mJy beam$^{-1}$. These observations complement the ATLAS observations, since they are wider, shallower and cover higher frequencies, targeting stronger, rarer and presumably younger AGN.

\begin{table*}
\caption{A summary of the ATCA and LBA observations (see section~\ref{obs}) undertaken in this project, sorted by date.}
\begin{tabular}{clcccrclc}
Sample & Telescope & Project Number & Date & Array Configuration & Frequency & Time & Field & $N_{\rm sources}$\\
\hline
(1) & ATCA & C2730 & Dec 2012 & 6B & 5.5/9~GHz & 39 h & ELAIS-S1 & 49\\
(2) & ATCA & C2768 & Feb 2013 & 6A & 5.5/9~GHz & 12 h & SMC & 72\\
(3) & LBA & V506a & Dec 2013 & At-Ak-Pa-Ho-Cd & 1.6~GHz & 10 h & ELAIS-S1 & 5\\
(4) & LBA & V506b & Feb 2014 & At-Ak-Pa-Ho-Cd-Mp & 1.6~GHz & 10 h & CDFS & 3\\
(5) & ATCA & C2730 & Apr 2014 & 6A & 5.5/9~GHz & 11 h & CDFS & 22\\
\end{tabular}
\label{obs_table}
\end{table*}

\subsubsection{Sample Selection}

\label{selection}

We followed the basic strategy of selecting unresolved GPS and CSS candidates to be observed at higher resolution, and then selecting the most compact of these (which remained unresolved) to be observed with VLBI. Our GPS and CSS candidates were selected according to the following criteria:

\begin{enumerate}
\item Has a GPS radio spectrum with an identifiable peak; or
\item Has a CSS radio spectrum with $\alpha < -0.8$; and
\item Is unresolved at the highest available resolution.
\end{enumerate}

The first two criteria were based on visual inspection of plots of the radio spectrum (e.g. Figure~\ref{s150}), which used a least-squares power-law fit to the available flux densities (including our new observations where available $-$ see section~\ref{obs}), which were measured from beam-matched images where possible. The third criterion gave an upper limit on the angular size of the source, ensuring that higher-resolution observations would yield a high enough signal-to-noise.

The five samples we selected are summarised in Table~\ref{obs_table}. Sources from samples (1), (2) and (5) were selected to be unresolved at 1.4~GHz (i.e. $\lesssim 10$ arcsec) in ATLAS and the SMC. Sources in sample (5) were selected to be outside the ECDFS, which had already been observed. Sources from sample (3) were selected to be unresolved at 9~GHz from sample (1) (i.e. $< 1$ arcsec). Sample (5) had not been observed when sample (4) was selected, and therefore these sources were selected slightly differently. One source from sample (4) was selected to be unresolved at 5.5~GHz in the ECDFS (i.e. $< 5$ arcsec). The other two sources from sample (4) were outside the 0.25~deg$^2$ of the ECDFS, and since their redshifts were known, the expected angular size (which we used to essentially replace the third criterion) was estimated based on the linear size derived from equation~\ref{linear_size_turnover}, which used a turnover frequency estimated from the radio spectrum. Sources in sample (4) were also selected to be not observed or detected with VLBI by \citet{2011A&A...526A..74M}.

In the SMC, since our resulting targets were relatively bright, we were able to select a handful of additional targets without losing much sensitivity across the sample. Therefore, we also selected sources that were resolved at the $\sim$10$\arcsec$ resolution, to be used as a reference sample of more extended RGs, which are presumably older FR I and FR II galaxies. 

The resulting sample consisted of 144 GPS and CSS candidates and RGs selected from ATLAS and the SMC, including the ECDFS source that was unique to sample (4).

\subsubsection{Observations}

\label{obs}

We undertook high-resolution observations using the new 4~cm receiver on the Australia Telescope Compact Array (ATCA $-$ proposal ID: C2730 and C2768) of the faintest GPS/CSS sample to date (samples (1), (2) and (5) from Table~\ref{obs_table}). These observations made use of the ATCA in 6A and 6B configurations, which used a 6 km baseline and gave a resolution at 9 GHz of $\sim$1$\arcsec$. Our data consisted of 2 GHz Compact Array Broadband Backend \citep[CABB;][]{CABB} observations at 5.5 and 9.0~GHz, centred on the selected sources. We observed the ATLAS targets down to varying r.m.s. levels between $\sim$10$-100~\mu$Jy beam$^{-1}$, and the SMC targets down to varying r.m.s. levels between $\sim$50 $-150~\mu$Jy beam$^{-1}$, depending on the strength of each source. The on-source time we scheduled for each pointing was chosen based on achieving a target signal-to-noise ratio (SNR) of 5$\sigma$ in each beam at 9~GHz. We expected the CSS sources to have a discontinuous jet morphology spanning $< 10$~kpc, which for a typical redshift of 0.5, gives an angular size of $\lesssim 3$ arcsec, spanning $\lesssim 3$ beams at 9~GHz. To estimate the SNR, we used the measured 8.64~GHz flux densities for SMC sources, and an estimated 9~GHz flux density extrapolated from the spectral index for ATLAS sources. 

After taking these new observations, we had radio flux density measurements at a minimum of 8 frequencies in ATLAS and 5 frequencies in the SMC, and many more if the source had a high enough SNR to enable the data to be broken up into smaller sub-bands. 

Additionally, we undertook Long Baseline Array (LBA $-$ proposal ID: V506) observations of the most compact sources selected from ATLAS (samples (3) and (4) from Table~\ref{obs_table}). These observations used ATCA, ASKAP, Parkes, Hobart and Ceduna (At-Ak-Pa-Ho-Cd), as well as Mopra (Mp) where available, giving resolutions as high as $\sim$15 mas. We plan to select further targets in ongoing VLBI observations, and present the results in Collier et al., in prep. 

Our GPS/CSS sources range in low to high redshift, small to large, and bright to faint AGN, with which we will construct our evolutionary sequence. This will allow us to directly test each hypothesis, in particular, what fraction of GPS/CSS sources grow to larger sizes. 

\begin{figure}
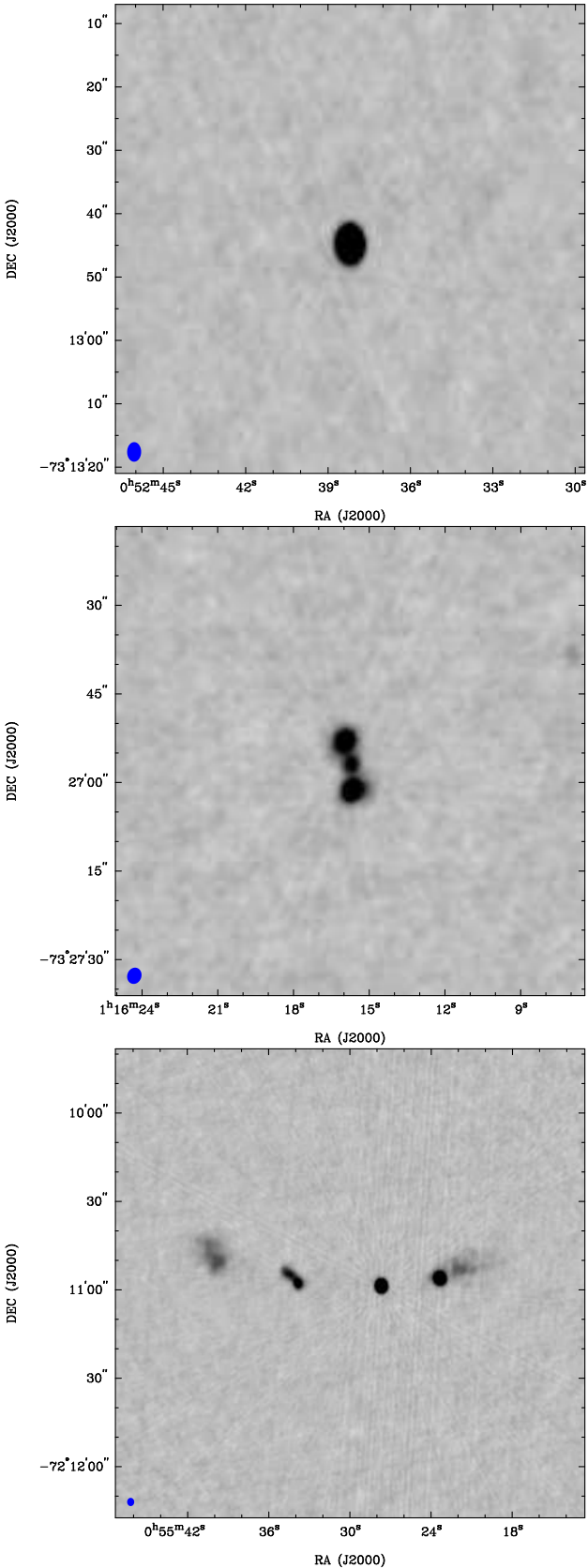

\includegraphics[width=0.43\textwidth,angle=-90]{fig2}
\includegraphics[width=0.43\textwidth,angle=-90]{fig3}
\includegraphics[width=0.43\textwidth,angle=-90]{fig4}
\caption{An example of an unresolved and resolved GPS/CSS candidate from the SMC, compared to one of the resolved RGs, all observed at 5.5~GHz and imaged with natural weighting during this project. The synthesised beam is shown by the blue ellipse in the bottom left corner.}
\label{RGs}
\end{figure}

\begin{figure}
\begin{center}
\includegraphics[width=0.48\textwidth]{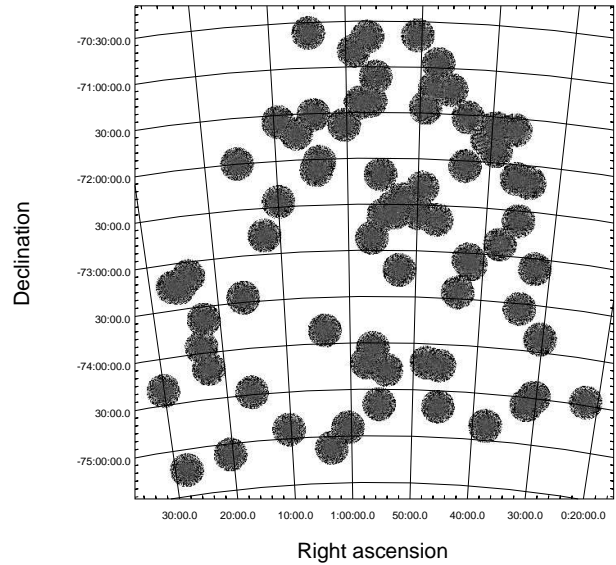}
\caption{A  72-pointing mosaic of the SMC at 5.5~GHz, imaged with natural weighting and consisting of over 2 billion pixels. This shows the region of mJy sky from which our GPS/CSS candidates were selected.}
\label{mosaic}
\end{center}
\end{figure}

\subsection{Early Results}

Fig.~\ref{RGs} shows a few examples of our sources from the SMC sample. Fig.~\ref{mosaic} shows a 45 181 by 46 597 pixel mosaic of all 72 SMC pointings. Within this field, using our previous classifications of the radio spectra made during the selection of the targets (see section~\ref{selection}), we classify GPS sources based on being unresolved in the 9~GHz ATCA observations. Using these limited frequencies, we are able to identify some GPS sources only tentatively, until we combine the new flux measurements from 5.5 and 9~GHz. Based on these initial classifications, we estimate that 2.5-12.5\% of the mJy sky behind the SMC consists of GPS sources, consistent with the \cite{ODea}  fraction of $\sim$10\% for all-sky Jy-level GPS sources. 

The full results will be published in a study of the SMC sample (Collier et al., 2015, in prep.) and the ATLAS sample (Collier et al., 2016, in prep.).

\section{Summary and Further Work}

\label{last_section}

We have observed the faintest GPS/CSS sample to date. So far, we've discovered that the mJy sky behind the SMC consists of 2.5-12.5\% GPS sources. Our high-resolution observations have allowed us to resolve the jets of many sources. The age derived from the jet sizes will be compared to other multi-wavelength age tracers. Using these, we will place GPS/CSS sources into an evolutionary sequence, along with other young AGN candidates, including IFRSs, which seem to be a related class of object, dominated by GPS/CSS sources at high-$z$. We will test each hypothesis for the nature of GPS/CSS sources, in particular, what fraction of them are young and evolving.

Above all, we will discover the properties of faint GPS/ CSS sources. The results will be presented in Collier et al. (2015) and Collier et al. (2016).

 \bibliographystyle{an}
 \bibliography{an-refs}

\begin{thebibliography}{46}
\expandafter\ifx\csname natexlab\endcsname\relax\def\natexlab#1{#1}\fi

\bibitem[{{Alexander}(2000)}]{2000MNRAS.319....8A}
{Alexander}, P. 2000, \mnras, 319, 8

\bibitem[{{Antonuccio-Delogu} \& {Silk}(2008)}]{2008MNRAS.389.1750A}
{Antonuccio-Delogu}, V. \& {Silk}, J. 2008, \mnras, 389, 1750

\bibitem[{{Baum} {et~al.}(1990){Baum}, {O'Dea}, {Murphy}, \& {de
  Bruyn}}]{1990A&A...232...19B}
{Baum}, S.~A., {O'Dea}, C.~P., {Murphy}, D.~W., \& {de Bruyn}, A.~G. 1990,
  \aap, 232, 19

\bibitem[{{Best} {et~al.}(2006){Best}, {Kaiser}, {Heckman}, \&
  {Kauffmann}}]{Best2006}
{Best}, P.~N., {Kaiser}, C.~R., {Heckman}, T.~M., \& {Kauffmann}, G. 2006,
  \mnras, 368, L67

\bibitem[{{Chiaberge} {et~al.}(2015){Chiaberge}, {Gilli}, {Lotz}, \&
  {Norman}}]{2015ApJ...806..147C}
{Chiaberge}, M., {Gilli}, R., {Lotz}, J.~M., \& {Norman}, C. 2015, \apj, 806,
  147

\bibitem[{{Collier} {et~al.}(2014){Collier}, {Banfield}, {Norris},
  {Schnitzeler}, {Kimball}, {Filipovi{\'c}}, {Jarrett}, {Lonsdale}, \&
  {Tothill}}]{Collier_2014}
{Collier}, J.~D., {Banfield}, J.~K., {Norris}, R.~P., {et~al.} 2014, \mnras,
  439, 545

\bibitem[{{Crawford} {et~al.}(2011){Crawford}, {Filipovic}, {de Horta}, {Wong},
  {Tothill}, {Draskovic}, {Collier}, \& {Galvin}}]{2011SerAJ.183...95C}
{Crawford}, E.~J., {Filipovic}, M.~D., {de Horta}, A.~Y., {et~al.} 2011,
  Serbian Astronomical Journal, 183, 95

\bibitem[{{Croton} {et~al.}(2006){Croton}, {Springel}, {White}, {De Lucia},
  {Frenk}, {Gao}, {Jenkins}, {Kauffmann}, {Navarro}, \& {Yoshida}}]{Croton2006}
{Croton}, D.~J., {Springel}, V., {White}, S.~D.~M., {et~al.} 2006, \mnras, 365,
  11

\bibitem[{{Fanaroff} \& {Riley}(1974)}]{1974MNRAS.167P..31F}
{Fanaroff}, B.~L. \& {Riley}, J.~M. 1974, \mnras, 167, 31P

\bibitem[{{Fanti}(2009{\natexlab{a}})}]{2009AN....330..120F}
{Fanti}, C. 2009{\natexlab{a}}, Astronomische Nachrichten, 330, 120

\bibitem[{{Fanti} {et~al.}(1995){Fanti}, {Fanti}, {Dallacasa}, {Schilizzi},
  {Spencer}, \& {Stanghellini}}]{1995A&A...302..317F}
{Fanti}, C., {Fanti}, R., {Dallacasa}, D., {et~al.} 1995, \aap, 302, 317

\bibitem[{{Fanti}(2009{\natexlab{b}})}]{2009AN....330..303F}
{Fanti}, R. 2009{\natexlab{b}}, Astronomische Nachrichten, 330, 303

\bibitem[{{Filipovi{\'c}} {et~al.}(2002){Filipovi{\'c}}, {Bohlsen}, {Reid},
  {Staveley-Smith}, {Jones}, {Nohejl}, \& {Goldstein}}]{2002MNRAS.335.1085F}
{Filipovi{\'c}}, M.~D., {Bohlsen}, T., {Reid}, W., {et~al.} 2002, \mnras, 335,
  1085

\bibitem[{{Hancock} {et~al.}(2010){Hancock}, {Sadler}, {Mahony}, \&
  {Ricci}}]{2010MNRAS.408.1187H}
{Hancock}, P.~J., {Sadler}, E.~M., {Mahony}, E.~K., \& {Ricci}, R. 2010,
  \mnras, 408, 1187

\bibitem[{{Hardcastle} {et~al.}(2007){Hardcastle}, {Evans}, \&
  {Croston}}]{Hardcastle2007}
{Hardcastle}, M.~J., {Evans}, D.~A., \& {Croston}, J.~H. 2007, \mnras, 376,
  1849

\bibitem[{{Herzog} {et~al.}(2014){Herzog}, {Middelberg}, {Norris}, {Sharp},
  {Spitler}, \& {Parker}}]{Andreas_IFRS}
{Herzog}, A., {Middelberg}, E., {Norris}, R.~P., {et~al.} 2014, \aap, 567, A104

\bibitem[{{Hopkins} {et~al.}(2008){Hopkins}, {Hernquist}, {Cox}, \& {Kere{\v
  s}}}]{2008ApJS..175..356H}
{Hopkins}, P.~F., {Hernquist}, L., {Cox}, T.~J., \& {Kere{\v s}}, D. 2008,
  \apjs, 175, 356

\bibitem[{{Huynh} {et~al.}(2012){Huynh}, {Hopkins}, {Lenc}, {Mao},
  {Middelberg}, {Norris}, \& {Randall}}]{2012MNRAS.426.2342H}
{Huynh}, M.~T., {Hopkins}, A.~M., {Lenc}, E., {et~al.} 2012, \mnras, 426, 2342

\bibitem[{{Koz{\l}owski} {et~al.}(2011){Koz{\l}owski}, {Kochanek}, \&
  {Udalski}}]{Kozlowski_2011}
{Koz{\l}owski}, S., {Kochanek}, C.~S., \& {Udalski}, A. 2011, \apjs, 194, 22

\bibitem[{{Koz{\l}owski} {et~al.}(2013){Koz{\l}owski}, {Onken}, {Kochanek},
  {Udalski}, {Szyma{\'n}ski}, {Kubiak}, {Pietrzy{\'n}ski}, {Soszy{\'n}ski},
  {Wyrzykowski}, {Ulaczyk}, {Poleski}, {Pietrukowicz}, {Skowron}, {OGLE
  Collaboration}, {Meixner}, \& {Bonanos}}]{Kozlowski_2013}
{Koz{\l}owski}, S., {Onken}, C.~A., {Kochanek}, C.~S., {et~al.} 2013, \apj,
  775, 92

\bibitem[{{Malbon} {et~al.}(2007){Malbon}, {Baugh}, {Frenk}, \&
  {Lacey}}]{2007MNRAS.382.1394M}
{Malbon}, R.~K., {Baugh}, C.~M., {Frenk}, C.~S., \& {Lacey}, C.~G. 2007,
  \mnras, 382, 1394

\bibitem[{{Mauch} {et~al.}(2003){Mauch}, {Murphy}, {Buttery}, {Curran},
  {Hunstead}, {Piestrzynski}, {Robertson}, \& {Sadler}}]{2003MNRAS.342.1117M}
{Mauch}, T., {Murphy}, T., {Buttery}, H.~J., {et~al.} 2003, \mnras, 342, 1117

\bibitem[{{Middelberg} {et~al.}(2011){Middelberg}, {Deller}, {Morgan},
  {Rottmann}, {Alef}, {Tingay}, {Norris}, {Bach}, {Brisken}, \&
  {Lenc}}]{2011A&A...526A..74M}
{Middelberg}, E., {Deller}, A., {Morgan}, J., {et~al.} 2011, \aap, 526, A74

\bibitem[{{Middelberg} {et~al.}(2008){Middelberg}, {Norris}, {Cornwell},
  {Voronkov}, {Siana}, {Boyle}, {Ciliegi}, {Jackson}, {Huynh}, {Berta},
  {Rubele}, {Lonsdale}, {Ivison}, \& {Smail}}]{ATLAS_ELAIS}
{Middelberg}, E., {Norris}, R.~P., {Cornwell}, T.~J., {et~al.} 2008, \aj, 135,
  1276

\bibitem[{{Murgia}(2003)}]{2003PASA...20...19M}
{Murgia}, M. 2003, \pasa, 20, 19

\bibitem[{{Norris} {et~al.}(2006){Norris}, {Afonso}, {Appleton}, {Boyle},
  {Ciliegi}, {Croom}, {Huynh}, {Jackson}, {Koekemoer}, {Lonsdale},
  {Middelberg}, {Mobasher}, {Oliver}, {Polletta}, {Siana}, {Smail}, \&
  {Voronkov}}]{ATLAS_CDFS}
{Norris}, R.~P., {Afonso}, J., {Appleton}, P.~N., {et~al.} 2006, \aj, 132, 2409

\bibitem[{{Norris} {et~al.}(2011{\natexlab{a}}){Norris}, {Afonso}, {Cava},
  {Farrah}, {Huynh}, {Ivison}, {Jarvis}, {Lacy}, {Mao}, {Maraston}, {Mauduit},
  {Middelberg}, {Oliver}, {Seymour}, \& {Surace}}]{Norris2011}
{Norris}, R.~P., {Afonso}, J., {Cava}, A., {et~al.} 2011{\natexlab{a}}, \apj,
  736, 55

\bibitem[{{Norris} {et~al.}(2011{\natexlab{b}}){Norris}, {Hopkins}, \& {et
  al.}}]{EMU}
{Norris}, R.~P., {Hopkins}, A.~M., \& {et al.} 2011{\natexlab{b}}, \pasa, 28,
  215

\bibitem[{{Norris} {et~al.}(2012){Norris}, {Lenc}, {Roy}, \&
  {Spoon}}]{Norris12}
{Norris}, R.~P., {Lenc}, E., {Roy}, A.~L., \& {Spoon}, H. 2012, \mnras, 422,
  1453

\bibitem[{{O'Dea}(1998)}]{ODea}
{O'Dea}, C.~P. 1998, \pasp, 110, 493

\bibitem[{{O'Dea} \& {Baum}(1997)}]{1997AJ....113..148O}
{O'Dea}, C.~P. \& {Baum}, S.~A. 1997, \aj, 113, 148

\bibitem[{{Orienti} {et~al.}(2010){Orienti}, {Murgia}, \&
  {Dallacasa}}]{2010MNRAS.402.1892O}
{Orienti}, M., {Murgia}, M., \& {Dallacasa}, D. 2010, \mnras, 402, 1892

\bibitem[{{Polatidis} \& {Conway}(2003)}]{2003PASA...20...69P}
{Polatidis}, A.~G. \& {Conway}, J.~E. 2003, \pasa, 20, 69

\bibitem[{{Randall} {et~al.}(2011){Randall}, {Hopkins}, {Norris}, \&
  {Edwards}}]{2011MNRAS.416.1135R}
{Randall}, K.~E., {Hopkins}, A.~M., {Norris}, R.~P., \& {Edwards}, P.~G. 2011,
  \mnras, 416, 1135

\bibitem[{{Randall} {et~al.}(2012){Randall}, {Hopkins}, {Norris}, {Zinn},
  {Middelberg}, {Mao}, \& {Sharp}}]{2012MNRAS.421.1644R}
{Randall}, K.~E., {Hopkins}, A.~M., {Norris}, R.~P., {et~al.} 2012, \mnras,
  421, 1644

\bibitem[{{Shabala} {et~al.}(2008){Shabala}, {Ash}, {Alexander}, \&
  {Riley}}]{Stas_2008}
{Shabala}, S.~S., {Ash}, S., {Alexander}, P., \& {Riley}, J.~M. 2008, \mnras,
  388, 625

\bibitem[{{Shabala} {et~al.}(2011){Shabala}, {Kaviraj}, \&
  {Silk}}]{2011MNRAS.413.2815S}
{Shabala}, S.~S., {Kaviraj}, S., \& {Silk}, J. 2011, \mnras, 413, 2815

\bibitem[{{Shabala} {et~al.}(2012){Shabala}, {Ting}, {Kaviraj}, {Lintott},
  {Crockett}, {Silk}, {Sarzi}, {Schawinski}, {Bamford}, \&
  {Edmondson}}]{2012MNRAS.423...59S}
{Shabala}, S.~S., {Ting}, Y.-S., {Kaviraj}, S., {et~al.} 2012, \mnras, 423, 59

\bibitem[{{Shulevski} {et~al.}(2012){Shulevski}, {Morganti}, {Oosterloo}, \&
  {Struve}}]{2012A&A...545A..91S}
{Shulevski}, A., {Morganti}, R., {Oosterloo}, T., \& {Struve}, C. 2012, \aap,
  545, A91

\bibitem[{{Sirothia} {et~al.}(2009){Sirothia}, {Dennefeld}, {Saikia}, {Dole},
  {Ricquebourg}, \& {Roland}}]{2009MNRAS.395..269S}
{Sirothia}, S.~K., {Dennefeld}, M., {Saikia}, D.~J., {et~al.} 2009, \mnras,
  395, 269

\bibitem[{{Tingay} {et~al.}(2015){Tingay}, {Macquart}, {Collier}, {Rees},
  {Callingham}, {Stevens}, {Carretti}, {Wayth}, {Wong}, {Trott}, {McKinley},
  {Bernardi}, {Bowman}, {Briggs}, {Cappallo}, {Corey}, {Deshpande}, {Emrich},
  {Gaensler}, {Goeke}, {Greenhill}, {Hazelton}, {Johnston-Hollitt}, {Kaplan},
  {Kasper}, {Kratzenberg}, {Lonsdale}, {Lynch}, {McWhirter}, {Mitchell},
  {Morales}, {Morgan}, {Oberoi}, {Ord}, {Prabu}, {Rogers}, {Roshi}, {Udaya
  Shankar}, {Srivani}, {Subrahmanyan}, {Waterson}, {Webster}, {Whitney},
  {Williams}, \& {Williams}}]{Tingay}
{Tingay}, S.~J., {Macquart}, J.-P., {Collier}, J.~D., {et~al.} 2015, \aj, 149,
  74

\bibitem[{{Tinti} \& {de Zotti}(2006)}]{2006A&A...445..889T}
{Tinti}, S. \& {de Zotti}, G. 2006, \aap, 445, 889

\bibitem[{{Turner} \& {Shabala}(2015)}]{Ross_2015}
{Turner}, R.~J. \& {Shabala}, S.~S. 2015, \apj, 806, 59

\bibitem[{{van Breugel} {et~al.}(1984){van Breugel}, {Miley}, \&
  {Heckman}}]{1984AJ.....89....5V}
{van Breugel}, W., {Miley}, G., \& {Heckman}, T. 1984, \aj, 89, 5

\bibitem[{{Wilson} {et~al.}(2011){Wilson}, {Ferris}, \& {et al.}}]{CABB}
{Wilson}, W.~E., {Ferris}, R.~H., \& {et al.} 2011, \mnras, 416, 832

\bibitem[{{Wong} {et~al.}(2011){Wong}, {Filipovic}, {Crawford}, {de Horta},
  {Galvin}, {Draskovic}, \& {Payne}}]{2011SerAJ.182...43W}
{Wong}, G.~F., {Filipovic}, M.~D., {Crawford}, E.~J., {et~al.} 2011, Serbian
  Astronomical Journal, 182, 43

\end{thebibliography}

\label{lastpage}
\end{document}